\documentclass[12pt]{article}
\usepackage{latexsym,amsmath,amsfonts,natbib,bm}
\usepackage{bbm}
\usepackage{amssymb,amsthm,dsfont}
\usepackage{graphicx}
\usepackage{mathtools}
\usepackage{setspace}
\usepackage{appendix}
\usepackage{comment,verbatim}
\usepackage{microtype}
\usepackage{xcolor}
\usepackage{multirow}
\usepackage{subcaption}
\RequirePackage{hyperref}

\DisableLigatures[f]{encoding = *, family = * }
\numberwithin{equation}{section}
\numberwithin{figure}{section}
\numberwithin{table}{section}

\def\a{\alpha}
\def\b{\beta}
\def\d{\delta}
\def\g{\gamma}
\def\eps{\epsilon}
\def\N{\mathcal{N}}

\def\s{\sigma}
\def\bb{\boldsymbol \b}
\def\bxi{\bm{\xi}}

\def\bmu{\boldsymbol \mu}
\def\Sig{\boldsymbol \Sigma}
\def\P{\mathbb{P}}

\def\min{\text{min}}
\def\t{\theta}
\def\T{\Theta}
\def\l{\lambda}
\def\E{\mathbb{E}}

\makeatletter
\def\@seccntformat#1{\@ifundefined{#1@cntformat}%
   {\csname the#1\endcsname\quad}  
   {\csname #1@cntformat\endcsname}
}
\let\oldappendix\appendix 
\renewcommand\appendix{%
    \oldappendix
    \newcommand{\section@cntformat}{\appendixname~\thesection\quad}
}
\makeatother

\title{\bf Multiclass classification of growth curves using random change points and heterogeneous random effects}
\author{Vincent Chin\footnote{Communicating author: \href{mailto:vincent.chin@student.unsw.edu.au}{\tt vincent.chin@student.unsw.edu.au}} 
\thanks{School of Mathematics and Statistics, University of New South Wales, Sydney, Australia.} ,
Jarod Y. L. Lee\thanks{School of Mathematical and Physical Sciences, University of Technology Sydney, Sydney, Australia.} ,
Louise M. Ryan\footnotemark[3] \thanks{Harvard T.H. Chan School of Public Health, Harvard University, Cambridge, U.S.A.} ,
\\ Robert Kohn\thanks{School of Economics, Australian School of Business, University of New South Wales, Sydney, Australia.} { and} 
Scott A. Sisson\footnotemark[2]}
\date{}

\begin{document}
\maketitle
\begin{abstract}
Faltering growth among children is a nutritional problem prevalent in low to medium income countries; it is generally defined as a slower rate of growth compared to a reference healthy population of the same age and gender. As faltering is closely associated with reduced physical, intellectual and economic productivity potential, it is important to identify faltered children and be able to characterise different growth patterns so that targeted treatments can be designed and administered. We introduce a multiclass classification model for growth trajectory that flexibly extends a current classification approach called the broken stick model, which is a piecewise linear model with breaks at fixed knot locations. Heterogeneity in growth patterns among children is captured using mixture distributed random effects, whereby the mixture components determine the classification of children into subgroups. The mixture distribution is modelled using a Dirichlet process prior, which avoids the need to choose the ``true" number of mixture components, and allows this to be driven by the complexity of the data. Because children have individual differences in the onset of growth stages, we introduce child-specific random change points. Simulation results show that the random change point model outperforms the broken stick model because it has fewer restrictions on knot locations. We illustrate our model on a longitudinal birth cohort from the Healthy Birth, Growth and Development knowledge integration project funded by the Bill and Melinda Gates Foundation. Analysis reveals 9 subgroups of children within the population which exhibit varying faltering trends between birth and age one.
\end{abstract}
\begin{flushleft}
\end{flushleft}
\doublespacing

\newpage
\section{Introduction}
According to the latest joint malnutrition estimates by the \cite*{unicef2018levels}, it is estimated that in 2017, stunted growth is prevalent in 22.2\% of the global population under the age of 5, or over 150 million children worldwide. This is particularly serious in low to medium income countries where the rate of stunting is 35.0\%. A major contributor to stunted growth is prolonged faltering, defined as a slower rate of growth compared to a reference healthy population of the same age and gender, which comes with adverse consequences such as increased susceptibility to diarrhoea and respiratory infections \citep{kossmann2000undernutrition}, abnormal neurointegrative development \citep{benitez1999dendritic} and capital loss to the labour market \citep{hoddinott2013economic}. Therefore, it is imperative to take early preventive measures to minimise these impacts. In order to implement preventive measures, faltered children must first be identified in the population. It is additionally important to distinguish between different growth patterns, as each type represents a particular growth behaviour and so merits a different response. For example, children who caught up on growth after having faltered may have benefited from the intake of better diets or nutritional supplements. Such  strategies can then be extended to other children in the cohort to improve their growth.

There are various statistical approaches for modelling growth curves. These can be broadly classified into principal component-based and regression-based methods. Principal component-based methods, such as the FACE algorithm \citep{xiao2016fast}, originate from functional data analysis \citep{ramsay2005functional} whereby inferences on curves are conducted by discretising the curves to estimate a covariance function on which functional principal component analysis is then performed. The underlying structures in the curves are then given by the resulting eigenfunctions weighted by the associated eigenvalues. However, functional methods are not suitable for sparse observations. This has made regression-based methods more popular in the literature. Common regression models include the linear mixed model \citep{lindstrom1988newton}, and the broken stick model with random effects \citep{de2010terneuzen}, which is a piecewise linear model with breaks at the knots. \cite{anderson2018comparing} compare the most common growth modelling approaches and find that the broken stick model, when used in conjunction with z-scores has superior performance in terms of out-of-sample prediction. Z-scores, such as the height-for-age z-score (HAZ), is a measure defined by the \cite{who2006length} that compares the anthropometric measurements of a child against a reference population of healthy children of the same age and gender.

Once a model is fitted, the next step is to classify the children into the different faltering trajectory groups based on velocities derived from the model. Existing methods can be broadly classified into threshold-based and model-based approaches. An example of threshold-based methods is given in \cite{leung2017conditional}, which suggests classifying those children with the lowest 10\% of values of random slope estimates extracted from a linear mixed model as being abnormal. Threshold-based methods are arbitrary and the resulting classifications are not comparable between different populations. In an attempt to overcome this caveat, \cite{lee2018detecting} proposed fitting a two-component mixture distribution based on minimum velocities extracted for each child from a broken stick model with random slopes. This is inefficient as information from a vector of growth velocities for each child is summarised into a scalar, and so it is then difficult to identify different types of growth trajectory. Furthermore, their adopted modelling assumptions on the random slopes appear to be somewhat contradictory between the regression model and the classifier.

Our article flexibly extends the broken stick model to introduce an approach which incorporates the classifier within the regression model. This allows the classification of growth curves into different patterns based on the vectors of velocities to be achieved within a single model. In order to capture heterogeneity in the growth velocity between children, we extend the broken stick model to allow for mixture distributed random slopes. Classification of an individual child's growth profile is then determined by the component of the mixture distribution from which the vector of velocities is generated. We note that the concept of using mixture distributed random effects within regression models for classification is not new. In the context of linear mixed models, \cite{verbeke1996linear} identify slow growers in a population of schoolgirls while \cite{xu2001random} classify treatment response of patients in clinical trials. Both articles use a finite mixture distribution, which requires specifying the number of components in advance. We adopt a different strategy: modelling the distribution of growth velocity non-parametrically within a Bayesian framework using the Dirichlet process (DP) prior. The DP prior adapts the complexity of the model to the amount of data available without requiring an a priori choice of the number of mixture components, which is often unknown in practical applications.

A further contribution of this paper is to introduce random change points for the knots into the broken stick model, rather than their locations being arbitrarily fixed. These change points are modelled as random effects so that the difference in the timing of growth phases between children can be accommodated within the model and the classification process. Probabilistic inference for these change points is straightforward and can be implemented using Markov chain Monte Carlo (MCMC) algorithms. Classification of each child's growth trajectory relies on the posterior distribution, which for mixture models is known to suffer from the label switching problem \citep{jasra2005markov}. Additionally, the number of components in a DP mixture model is variable. We overcome these issues by implementing the posterior expected adjusted Rand (PEAR) method proposed in \cite{fritsch2009improved}, which is based on Bayesian decision theory, to classify each child.

Our simulation studies demonstrate the superior performance of the random change points model compared to the broken stick model with fixed knots. For the latter, due to its limited flexibility in capturing change point heterogeneity, the number of clusters tends to be overestimated in our analyses because of the biased estimation of growth velocity. We apply the proposed methodology by analysing the growth profiles of a birth cohort of 373 children from Vellore, India, over their first year of life. The results suggest that there are 9 subgroups of children present in the population, with a majority exhibiting improved growth followed by a faltering trend.

\newpage
The paper is organised as follows: Section~\ref{sec:methods} describes the extension of the broken stick model to mixture distributed random effects using the DP prior. It also introduces individual-specific random change points to the model, and provides implementation details. Section~\ref{sec:simulation} investigates the performance of the proposed model via simulation studies. Section~\ref{sec:vellore} provides an analysis of the Vellore growth curve dataset, and Section~\ref{sec:conclusion} concludes.

\section{Methods} \label{sec:methods}

\subsection{A broken stick model with mixture distributed random slopes} \label{subsec:broken stick model}

A popular method for modelling longitudinal growth data in the epidemiological literature is the broken stick model. This may be defined as
\begin{equation}
z_{ij} = \a_i + \b_{0i} (t_{ij} - (t_{ij} - \xi_1)_+) + \b_{Ki} (t_{ij} - \xi_K)_+ + \sum_{k=1}^{K-1} \b_{ki} ((t_{ij} - \xi_k)_+ - (t_{ij} - \xi_{k+1})_+) + \eps_{ij},
\label{eqn:broken-stick}
\end{equation}
\begin{equation}
\a_i \sim \N(\mu_\a, \s^2_\a), \quad \eps_{ij} \sim \N(0, \s^2_\eps),
\label{eqn:random effects and errors}
\end{equation}
for $i=1, \dotsc, N, j=1, \dotsc, J_i$, where $z_{ij} \in \mathbb{R}$ denotes the height-for-age z-score (HAZ) for child $i$ on the $j$-th measurement occasion at age $t_{ij}$, $(x)_+=\max\{0,x\}$ is the positive part of $x$ and $\bxi=(\xi_1, \dotsc, \xi_K)^\top$ is an ordered vector of $K$ predetermined knots, or change points, such that $\xi_1 < \cdots < \xi_K$. The individual-specific random intercept $\a_i$ and error $\eps_{ij}$ are both assumed to be independent and normally distributed with parameter vectors given by $(\mu_\a, \s^2_\a)$ and $(0, \s^2_\eps)$ respectively. The child-specific and time invariant $\a_i$ controls for heterogeneity in the HAZ at birth, centred around the population mean $\mu_\a$, and is assumed to be uncorrelated with the error term $\eps_{ij}$. The broken stick model fits $K+1$ piecewise linear segments with breaks at $\bxi$ to model the growth trajectory calibrated in terms of the HAZ. The formulation in (\ref{eqn:broken-stick}) enables an individual child's growth velocity to be obtained directly from the regression coefficients since $\b_{ki}$ represents the rate of change in the HAZ between years $\xi_k$ and $\xi_{k+1}$.

\newpage
We now consider distributional assumptions on the individual-specific growth velocity vector $\bb_i = (\b_{0i}, \dotsc \b_{Ki})^\top$. \cite{anderson2018comparing} and \cite{lee2018detecting} model $\bb_i$ as realisations from a multivariate $\N(\bmu_{\bb}, \Sig_{\bb})$ distribution with mean vector $\bmu_{\bb}$ and covariance matrix $\Sig_{\bb}$. This signifies a homogeneous population model where individual growth profiles largely follow the trend of a global trajectory, with the variability of deviation from this mean curve determined by $\Sig_{\bb}$. Under this assumption the growth rate is, on average, the same for all children in the population. However, this is rarely the case in practice. For example, \cite{goode2014family} find that higher socio-economic status has a positive impact on the HAZ through greater health consciousness and better household sanitation systems. Studies have also found evidence of correlation between growth velocity during childhood and biological factors such as maternal height \citep{ramakrishnan1999role}. Therefore, we alternatively consider a more structured normal mixture distribution
\begin{equation}
\bb_i \sim \sum_{g=1}^G \pi_g \N(\bmu_g, \Sig_g),
\label{eqn:finite mixture}
\end{equation}
for positive weights $\pi_g>0$ with $\sum_{g=1}^G\pi_g=1$, in order to accommodate more complex compositions in the population. Each mixture component in (\ref{eqn:finite mixture}) corresponds to a particular type of growth pattern, and each child belongs (probabilistically) to one of these $G$ subgroups. By clustering the children into different subgroups, subsequent analyses can then identify risk factors which cause the manifestation of certain growth behaviours.

Equation (\ref{eqn:finite mixture}) requires specifying the number of subgroups $G$, which is typically unknown a priori in practice. There is an extensive existing literature that discusses this technical difficulty. One common approach is to perform a likelihood ratio test \citep{titterington1985statistical}, but the asymptotic distribution of the test statistic under the null hypothesis is unknown \citep{ghosh1985on}, as opposed to the conventional $\chi^2$ distribution. \cite{verbeke1996linear} consider a goodness-of-fit test by comparing the probability distribution of random variables derived from linear combinations of the observations against a uniform distribution using the Kolmogorov-Smirnov test. From the Bayesian perspective, \cite{dasgupta1998detecting} use the Bayesian information criterion (BIC) approximation to the Bayes factor as a basis for the selection of $G$, from which there is strong evidence to prefer the model with a larger value of $G$ if the BIC value increases by more than 10 upon an increase of one additional mixture component. \cite{sugar2003finding} propose computing the average Mahalanobis distance between the observations and their respective subgroup means for a range of values $G$. They show theoretically that the ``true" value of $G$ contributes to the largest drop in the distance. More recently, \cite{fuquene2016choosing} develop a family of repulsive prior distributions to penalise recurring components so that each subgroup is well distinguished.  In the next section, we describe a Bayesian approach that incorporates the estimation of $G$ via a Dirichlet process prior.

\subsection{Bayesian non-parametric mixture modelling}

Choosing a suitable value for the number of components $G$ in a mixture distribution is a non-trivial problem. Most of the methods described in Section~\ref{subsec:broken stick model} are ad-hoc, requiring the need to fit multiple models of differing complexity, and selecting the ``best" model based on certain criteria. In order to circumvent the model selection procedure, we employ a Bayesian non-parametric approach to fit a model which allows its complexity to be completely data-driven. In general, such flexibility is achieved by assuming an infinite dimensional parameter space $\T$, on which a prior distribution is then developed. In our present context, subscribing to this framework leads to an infinite mixture model.

A well-defined prior on $\T$, which is widely used in applications of mixture modelling \citep{zhang2005probabilistic, da2007dirichlet} is the Dirichlet process (DP) prior, established in \cite{ferguson1973bayesian}. Let $\mathcal{DP}(\l, \mathcal{H}_0)$ denote a Dirichlet process with concentration parameter $\l > 0$ and base distribution $\mathcal{H}_0$. A realisation, $\mathcal{H}$, from this stochastic process is a discrete probability distribution,  taking the  form
\begin{equation}
    \mathcal{H} = \sum_{g=1}^\infty \pi_g \d_{\t_g},
    \label{eqn:DP realisation}
\end{equation}
where $\d_\t$ is a point mass located at $\t$, and $\t_g, g=1\,\ldots,\infty$, are independent random samples drawn from $\mathcal{H}_0$. The infinite sequence of weights $\{\pi_g\}_{g=1}^\infty$ is typically constructed using the stick-breaking process \citep{sethuraman1994constructive} whereby metaphorically, a stick, initially of unit length, is repeatedly broken at a random lengths, as determined by a Beta random variable $\g$. In such a manner, the weights are constructed as
\begin{equation}
    \pi_g = \g_g \prod_{h=1}^{g-1} (1-\g_h), \quad \g_g \sim \textrm{Beta}(1, \l).
    \label{eqn:stick breaking}
\end{equation}
For any measurable set $A$ of $\T$, $\E[\mathcal{H}(A)] = \mathcal{H}_0(A)$, so that the prior for $\theta$ is centred on  $\mathcal{H}_0$. Loosely, this means that $\mathcal{H}_0$ can be considered as the average prior distribution for $\theta$. The concentration parameter $\l$ controls the variability of $\mathcal{H}$ around $\mathcal{H}_0$. In the limit as $\l \rightarrow \infty$, $\mathcal{H}$ converges to $\mathcal{H}_0$ pointwise, whereas $\mathcal{H}$ collapses to a point distribution as $\l \rightarrow 0$. Because of the discrete nature of $\mathcal{H}$, shown in (\ref{eqn:DP realisation}), samples generated from $\mathcal{H}$ have a positive probability of being identical. For example, the probability of generating exactly $\theta_g\sim\mathcal{H}$ is $\pi_g$. This property of the DP prior makes it a popular and attractive choice in clustering problems when the number of clusters is unknown.

In the current setting, we consider a DP mixture model \citep{antoniak1974mixtures} for $\bb_i$ for which
\begin{equation}
\bb_i | (\bmu_i, \Sig_i) \sim \N(\bmu_i, \Sig_i), \quad (\bmu_i, \Sig_i) | \mathcal{H} \sim \mathcal{H}, \quad \mathcal{H} \sim \mathcal{DP}(\l, \mathcal{H}_0),
\label{eqn:DPM}
\end{equation}
for $i=1, \dotsc, N$, where $\phi_i = (\bmu_i, \Sig_i)$ are the parameters of a normal distribution specifying the mixture component from which the growth velocity $\bb_i$ of child $i$ is generated. Since the parameter of interest is a mean vector and covariance matrix pair, one common choice of $\mathcal{H}_0$ is the normal-inverse-Wishart distribution with parameters $(\bm{m}, c, \nu, \Psi)$, having density function
\begin{equation*}
    p(\phi) \propto |\Sig|^{-1/2} \exp\bigg(-\frac{c}{2}(\bmu-\bm{m})^\top \Sig^{-1} (\bmu-\bm{m})\bigg) \times |\Sig|^{-(\nu+K+2)/2}\exp\bigg(-\frac{1}{2}\textrm{tr}(\Psi \Sig^{-1})\bigg). 
\end{equation*}
Integrating out $\mathcal{H}$ from (\ref{eqn:DPM}), \cite{blackwell1973ferguson} show that the conditional prior distribution induced on $\phi_i$ follows a P{\'o}lya urn scheme, constructed as
\begin{equation}
\phi_i | \phi_1, \dotsc, \phi_{i-1} \sim \frac{1}{\l+i-1} \sum_{j=1}^{i-1} \d_{\phi_j} + \frac{\l}{\l+i-1} \mathcal{H}_0.
\label{eqn:polya urn}
\end{equation}
From (\ref{eqn:polya urn}), the generating mechanism of the first parameter $\phi_1$ involves drawing an independent sample from the base distribution $\mathcal{H}_0$. Subsequent samples, $\phi_i$, are then obtained by setting $\phi_i$ to be a random draw from the previous samples $\{\phi_1, \dotsc, \phi_{i-1}\}$ with probability proportional to $i-1$ (thereby directly introducing a clustering effect within the mixture model) or a new sample from $\mathcal{H}_0$ (i.e.~a new mixture component) with probability proportional to $\l$. Accordingly, the generated samples $\{\phi_1, \dotsc, \phi_N\}$ concentrate on a set of unique values $\{\t_1, \dotsc, \t_G\}$, with a larger value of $\l$ giving rise to a larger (random) value of $G$. In fact, \cite{teh2011dirichlet} show that for $N, \l \gg 0$, we have
\begin{equation*}
\E[G] \simeq \l \log \bigg( 1+\frac{N}{\l} \bigg),
\label{eqn:mean of G}
\end{equation*}
indicating that the mean of $G$ scales logarithmically with the size of the dataset, $N$. Note that the value of $G$ is bound above by $N$, and its (random) value is determined as part of the posterior inference.

\subsection{Knot locations as random effects} \label{subsec:knot}

So far, the knot location vector $\bxi$ has been treated as predetermined and fixed across all children in the population. However, this is unrealistic in the current context as individual children react differently to treatment interventions such as the administration of vitamins or to negative experiences such as infections, which will likely occur at individual-specific time points. The heterogeneity in the timing of such events is likely to cause individual trajectories to change course at different time points. Furthermore, erroneously fixing $\bxi$ as in the broken stick model, will result in a biased estimate of the growth velocity $\bb_i$ as the regression lines between two neighbouring segments are connected at the knot. This then affects the classification of each child because their growth patterns are summarised by $\bb_i$. Therefore, a sensible approach is to model the knot locations within the interval of $[0, T]$ as a child-specific ordered vector of knot random effects $\bxi_i=(\xi_{i1}, \dotsc, \xi_{iK})^\top$. We construct the prior distribution of each $\bxi_i$ as
\begin{equation}
p(\bxi_i) \propto  \prod_{k=1}^{K+1}(\xi_{ik} - \xi_{i,k-1}) \times \prod_{k=1}^K \mathds{1} \Bigg(\xi_{ik} \in \bigg(\frac{(k-1)T}{K}, \frac{kT}{K}\bigg) \Bigg),
\label{eqn:order statistics}
\end{equation}
for $i=1, \dotsc, N$, where $\xi_{i0} = 0$ and $\xi_{i,K+1}=T$ for convenience and $\mathds{1}(E)$ is an indicator function which takes value 1 if the event $E$ occurs and 0 otherwise. The first product term in (\ref{eqn:order statistics}) is the distribution of the even-numbered order statistics from $2K+1$ points uniformly distributed on $[0, T]$, as used in \cite{green1995reversible}, which probabilistically encourages consecutive knot points to be uniformly spaced. Although it usefully penalises short subintervals, it would still be possible for the knots $\bxi_i$ to be concentrated in regions where there is an abundance of informative data. As such, we additionally  impose a hard constraint via the second product term in (\ref{eqn:order statistics}), which ensures that there is exactly one knot within each of the $K$ subintervals of equal length on $[0, T]$ (see e.g.~\cite{fan2010bayesian} for a similar construction).

\subsection{Posterior inference and cluster analysis} \label{subsec:implementation}

Posterior simulation for the DP mixture model defined in (\ref{eqn:DPM}) is straightforward to implement using MCMC methods \citep{gelman2013bayesian}. However, naive sampling schemes based on the P{\'o}lya urn construction of the DP prior in (\ref{eqn:polya urn}) can be highly inefficient due to numerical approximations of high dimensional integrals when the dimension of $\bb_i$ is large. Let $\bm{s}=(s_1, \dotsc, s_N)^\top, s_i\in\{1,2,\ldots\}$ be the vector of cluster allocation variables determining which subgroup each child belongs to. Here we focus on the MCMC sampling of $\bm{s}$, the weight of each mixture component $\pi_g$, the concentration parameter $\l$ of the DP prior, and the knot random effects $\bxi_i$, as  MCMC updates for other model parameters are straightforward. We implement the slice sampler proposed by \cite{walker2007sampling} which is based on the stick-breaking representation. The slice sampling algorithm introduces auxiliary variables $u_i, i=1, \dotsc, N$, whose distribution conditional on the label $s_i$ is uniform on $[0, \pi_{s_i}]$. This parameter augmentation strategy gives the conditional posterior distribution $\P(s_i = g | \cdots)$ of $s_i$ as
\begin{equation}
\P(s_i = g | \cdots) = \frac{\exp(\ell_i(\t_g)) \mathds{1}(\pi_g > u_i)}{\sum_{h:\pi_h > u_i} \exp(\ell_i(\t_h))},
\label{eqn:multinomial sampling}
\end{equation}
where $\ell_i(\t_g)$ is the log-likelihood function for child $i$ under group $g$. Conditional on the other model parameters, equation (\ref{eqn:multinomial sampling}) indicates that the possible subgroups to which any child belongs are restricted to a finite set of components in the infinite dimensional parameter space $\T$ whose weights are greater than $u_i$. Given this, the probability of child $i$ belonging to any of these subgroups is then proportional to the appropriate likelihood term $\exp(\ell_i(\t_g))$ for each group.

Denoting the number of children in the $g$-th occupied mixture component by $N_g, g=1, \dotsc, G,$ the conditional posterior distribution of the weights can be shown to be Dirichlet distributed \citep{ge2015distributed}, i.e.
\begin{equation*}
    (\pi_1, \dotsc, \pi_G, \pi')|\cdots \sim \textrm{Dirichlet}(N_1, \dotsc, N_G, \l),
\end{equation*}
where $\pi'=1-\sum_{g=1}^G\pi_g$ is the weight on $\T' = \T \backslash \{\t_1, \dotsc, \t_G\}$. The stick-breaking process in (\ref{eqn:stick breaking}) is then applied to $\pi'$ until the length of the stick is less than $\textrm{min} \{u_1, \dotsc, u_N\}$. For each additional break of the stick with initial length $\pi'$, a new sample $\t'\in\T'$ is drawn from the base distribution, $\mathcal{H}_0$. The rationale behind this is to ensure that $\T'$ has zero probability of being sampled in (\ref{eqn:multinomial sampling}). The generation of additional empty mixture components and the removal of unoccupied components after sampling $\bm{s}$ changes the value of $G$ between MCMC iterations.

\cite{escobar1995bayesian} show that likelihood function of the hyperparameter $\l$ is given by
\begin{equation*}
L(\l | \cdots)\propto \l^G \frac{\Gamma(\l)}{\Gamma(\l+N)} \propto \l^G \int_0^1 c^{\l-1} (1-c)^{N-1} dc,
\label{eqn:likelihood concentration}
\end{equation*}
where $\Gamma$ denotes the gamma function and $c$ is a latent variable. Under a Gamma$(a_\l, b_\l)$ prior for $\l$, the MCMC update for $(\l, c)$ can be performed by Gibbs sampling, whereby $c$ is generated from a Beta$(\l, N)$ distribution and then $\l$ from a Gamma$(a_\l + G, b_\l - \log c)$ distribution.

The individual-specific knots, $\bxi_i,$ can be updated one knot component, $k=1,\ldots,K$, at a time using a Metropolis-Hastings update. Writing $\tilde{\bxi}_i^{(k)} = (\xi_{i1}, \dotsc, \xi_{i,k-1}, \tilde{\xi}_{ik}, \xi_{i,k+1}, \dotsc, \xi_{iK})^\top$ as the proposed vector of knot locations with $\tilde{\xi}_{ik}$ sampled uniformly from the subinterval $((k-1)T/K, kT/K),$ the probability of accepting the proposal is given by
\begin{equation*}
\min \Bigg\{1, \exp(\ell_i(\tilde{\bxi}_i^{(k)})- \ell_i(\bxi_i)) \times \frac{(\xi_{i,k+1} - \tilde{\xi}_{ik})(\tilde{\xi}_{ik} - \xi_{i,k-1})}{(\xi_{i,k+1} - \xi_{ik})(\xi_{ik} - \xi_{i,k-1})} \Bigg\},
\label{eqn:MH ratio}
\end{equation*}
where $\ell_i$ denotes the log-likelihood for child $i$. For improved efficiency, the update for $\bxi_i$ can be performed in parallel for each child.

Bayesian inference for mixture models often suffers from the label switching problem \citep{jasra2005markov} due to the invariance of the likelihood function for $\bb_i$ in (\ref{eqn:finite mixture}) to permutations of the labels of the mixture components. This makes identification of and inference for each component (i.e.~the clusters of individuals) challenging. Identification of each component is further complicated in a DP mixture model since $G$ is variable. \cite{fritsch2009improved} proposed addressing this issue using Bayesian decision theory, where the best decision rule (for component membership) satisfies certain optimality conditions. A popular measure used for comparing competing membership clusterings $\bm{s}$ and $\tilde{\bm{s}},$ with $G$ and $\tilde{G}$ clusters respectively, is the adjusted Rand index (ARI; \citealp{hubert1985comparing}) defined as
\begin{equation*}
ARI(\bm{s},\tilde{\bm{s}}) = \frac{\sum_{p=1}^G \sum_{q=1}^{\tilde{G}} \binom{N_{pq}}{2} - \sum_{p=1}^G \binom{N_{p+}}{2} \sum_{q=1}^{\tilde{G}}\binom{N_{+q}}{2} / \binom{N}{2}}{\frac{1}{2} \big( \sum_{p=1}^G \binom{N_{p+}}{2} + \sum_{q=1}^{\tilde{G}}\binom{N_{+q}}{2} \big) - \sum_{p=1}^G \binom{N_{p+}}{2} \sum_{q=1}^{\tilde{G}}\binom{N_{+q}}{2} / \binom{N}{2}}. 
\notag
\end{equation*}
Here $N_{pq}$ is the number of individuals in group $p$ of membership clustering $\bm{s}$ that are also in group $q$ of membership clustering $\tilde{\bm{s}}$, $N_{p+} = \sum_{q=1}^{\tilde{G}} N_{pq}, N_{+q} = \sum_{p=1}^G N_{pq}$ and $\binom{n}{2}=n(n-1)/2$ is the Binomial coefficient. The ARI measures the similarity between the two clusterings, mostly taking values between 0 (for completely random clustering) and 1 (for identical clusterings). Negative values of the ARI are possible but they have no substantive use. In the current context, the optimal clustering $\hat{\bm{s}}$ maximises the posterior expected adjusted Rand (PEAR) index. That is,
\begin{equation*}
    \hat{\bm{s}} = \underset{\tilde{\bm{s}}}{\arg\max}\, \E_{\bm{s}}[ARI(\bm{s},\tilde{\bm{s}})],
\end{equation*}
where the expectation $\mathbb{E}_{\bm{s}}$ is taken with respect to the posterior distribution of $\bm{s}$.

\section{Simulation study}
\label{sec:simulation}

\begin{figure}
\centering
\includegraphics[trim=3cm 2.9cm 3cm 3.1cm,clip,width=\columnwidth]{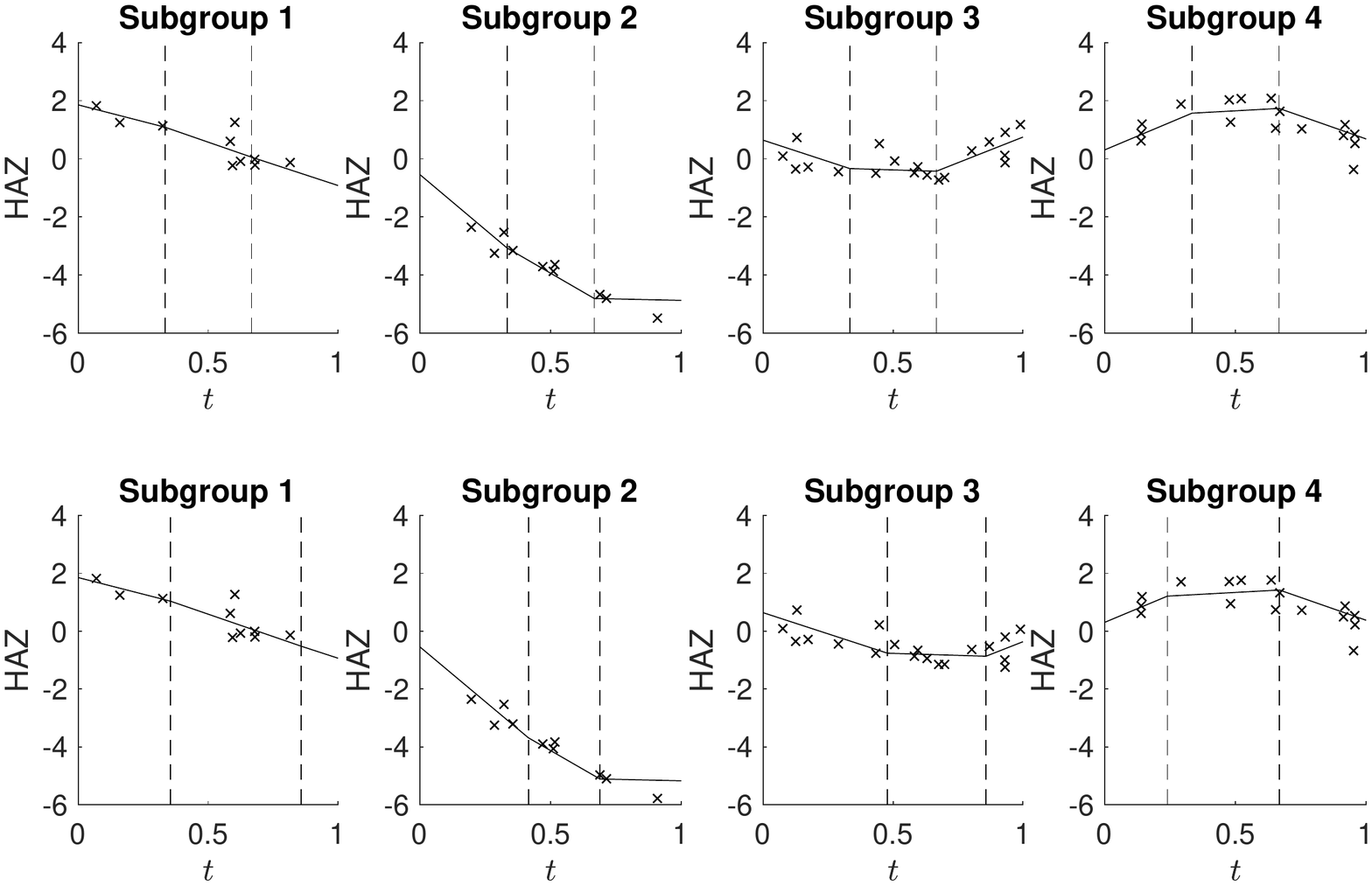}
\caption{\small{
HAZ score versus age (from birth until year 1) of representative simulated individuals from each of the four distinct growth trajectory groups (columns) under the broken stick model. Growth curve knot points are indicated by vertical dashed lines; the top panels showing equally spaced fixed knots ($D_{fixed}$) and the bottom panels showing the same individuals but with random knot points ($D_{random}$). The observed data ($\times$) is generated using the same random errors around each growth curve for each individual (top versus bottom panel in each column).
}}
\label{fig:trajectory simulated}
\end{figure}

We now examine how the above model and inferential procedure performs in a controlled setting. A population of $N=400$ children was generated under the broken stick model in (\ref{eqn:broken-stick}). The child-specific random intercepts $\a_i$ follow a $\N(0.75, 0.5)$ distribution and the error variance $\s^2_\eps$ is set as 0.15. The individual growth velocity vectors $\bb_i$ are generated from a normal mixture distribution with $G=4$ components with equal weights. The number of knots is specified as $K=2$ so that the growth trajectories are constructed from three piecewise linear segments. The mean velocities for each subgroup, $\bmu_g$, $g=1,\ldots, 4$, are given by 
\[
\begin{bmatrix}
\begin{array}{rrrr}
    \bmu_1 & \bmu_2 & \bmu_3 & \bmu_4
\end{array}
\end{bmatrix}
=
\begin{bmatrix}
\begin{array}{rrrr}
    -3.0 & -7.5 & -3.0 &  4.0 \\
    -3.0 & -5.0 & -1.0 &  1.0 \\
    -3.0 & { }0.0 &  3.0 & -3.0
\end{array}
\end{bmatrix},
\]
and the covariance matrix for each subgroup, $\Sig_g=0.2\bm{I}$, where $\bm{I}$ is the identity matrix. 
Using the same $\bb_i$, we construct two sets of data to examine different designs on the knot locations. The first dataset ($D_{fixed}$) has fixed and equally spaced knots at $t = \frac{1}{3}$ and $t = \frac{2}{3},$ while the second dataset ($D_{random}$) generates random knots for each child, with the first and second knots drawn uniformly from the intervals $[0, 0.5]$ and $[0.5, 1]$ respectively.
Figure~\ref{fig:trajectory simulated} illustrates growth profiles for one representative individual from each subgroup (columns) and also compares their differences between $D_{fixed}$ (top panels) and $D_{random}$ (bottom panels) for the same $\bb_i$. The first three subgroups exhibit a faltering pattern with different rates during the first two time periods. This faltering then either continues (subgroup 1), plateaus (subgroup 2) or growth improves (subgroup 3) in the third time period. In contrast, subgroup 4 is qualitatively different, whereby the children experience accelerated growth over time before a decline in the HAZ score is observed closer to age 1.

For each child, a random number (uniformly between 10 and 20) of HAZ observations was generated, with the measurement time of each observation being uniformly distributed between birth and age 1. To reduce variability in the comparison between $D_{fixed}$ and $D_{random}$, the two datasets were generated using the same number, measurement time of observations and random deviations around the two growth trajectories for each individual child. In this manner, the empirical residuals around each growth trajectory are identical between the two datasets for each child (see top versus bottom panels for each column in Figure~\ref{fig:trajectory simulated}).

\newpage
For inference we adopted weakly informative conjugate prior distributions. In particular we specified $\mu_\a \sim \N(0, 25)$, $\s_\a, \s_\eps \sim \textrm{half-Cauchy}(5)$, $\l \sim \textrm{Gamma}(2,4)$, and the base distribution $\mathcal{H}_0$ has a normal-inverse-Wishart distribution with parameters $(\bm{m}, c, \nu, \Psi)=(\bm{0}, 10^{-3}, K+2, \bm{I})$. Posterior sampling is achieved using MCMC following the details in Section \ref{subsec:implementation} with a chain of 100\,000 iterations, with the first 50\,000 iterations discarded as burn in and retaining every 20th of the remaining samples. We fit two model variants to each dataset: $M_{fixed}$ is the model with $K=2$ fixed and equally spaced knots at $t=\frac{1}{3}$ and $t=\frac{2}{3}$, whereas $M_{random}$ is the model where the two knot points are allowed to vary for each individual. Optimisation of PEAR to obtain $\hat{\bm{s}}$ was implemented using the \verb|mcclust| package in \verb|R|.

\begin{table}[b!]
 \centering
 \def\~{\hphantom{0}}
  \begin{tabular*}{\textwidth}{@{}c@{\extracolsep{\fill}}c@{\extracolsep{\fill}}c@{\extracolsep{\fill}}c@{\extracolsep{\fill}}c@{\extracolsep{\fill}}c@{\extracolsep{\fill}}c@{\extracolsep{\fill}}c@{\extracolsep{\fill}}c@{}}
  \hline \hline
Dataset & Model & $G_{min}$ & $G_{max}$ & $G_{mode}$ & $\hat{G}$ & $\E_{\bm{s}}\left[ARI(\hat{\bm{s}},\bm{s})\right]$ & $ARI(\hat{\bm{s}}, \bm{s}_{true})$ \\ \hline
\multirow{2}{*}{$D_{fixed}$} & $M_{fixed}$ & 4 & 6 & 4 & 4 & 0.9674 & 0.9734 \\
\cline{2-8}
& $M_{random}$ & 4 & 5 & 4 & 4 & 0.9096 & 0.9606 \\ \hline
\multirow{2}{*}{$D_{random}$} & $M_{fixed}$ & 6 & 9 & 7 & 7 & 0.7096 & 0.6102 \\
\cline{2-8}
& $M_{random}$ & 4 & 5 & 4 & 4 & 0.8245 & 0.7756 \\
\hline
\end{tabular*}
\caption{Performance summary when fitting fixed and random knot location models ($M_{fixed}$ and $M_{random}$) to fixed and random knot location datasets ($D_{fixed}$ and $D_{random}$). For each dataset/model pair, columns indicate minimum, maximum and mode of the posterior of the number of mixture components ($G_{min}$, $G_{max}$, $G_{mode}$); the number of groups $\hat{G}$ in the optimal clustering $\hat{s}$; the value of the posterior expectation $\mathbb{E}_{\bm{s}}\left[ARI(\hat{\bm{s}},\bm{s})\right]$) evaluated at $\hat{\bm{s}}$; and the ARI score comparing the estimated $\hat{\bm{s}}$ to the true group structure $\bm{s}_{true}$.}
\label{tab:results}
\end{table}
Table~\ref{tab:results} presents a summary of the performance of models $M_{fixed}$ and $M_{random}$ for both datasets, in terms of the final group classification outcome. For the fixed knot dataset $D_{fixed}$ both models perform similarly well by correctly identifying the true number of groups. This largely occurs as the $M_{fixed}$ model is contained within the $M_{random}$ model, and so the latter has the capacity to achieve the same performance as the former when the data  have the knot structure assumed in $M_{fixed}$. Of course, here model $M_{fixed}$ is slightly outperforming $M_{random}$ in terms of the $\E_{\bm{s}}\left[ARI(\hat{\bm{s}},\bm{s})\right]$ because the latter needs to estimate the knot locations based on a small number of observed datapoints, which introduces some variability into the final classification. As the number of observations per individual increases, we can expect these two models to perform similarly. Although model $M_{random}$ produces lower agreement with the true clustering $\bm{s}_{true}$, the realised classification obtained from the optimisation of PEAR is very much comparable to that of model $M_{fixed}$ as shown in Table~\ref{tab:PEVI}: most children are allocated to their respective true groups, with a $1.5\%$ misclassification rate.

\begin{table}[b!]
 \centering
 \def\~{\hphantom{0}}
 \begin{minipage}{\textwidth}
  \begin{tabular*}{\textwidth}{@{}c@{\extracolsep{\fill}}c@{\extracolsep{\fill}}c@{\extracolsep{\fill}}c@{\extracolsep{\fill}}c@{\extracolsep{\fill}}c@{\extracolsep{\fill}}c@{\extracolsep{\fill}}c@{\extracolsep{\fill}}c@{\extracolsep{\fill}}c@{\extracolsep{\fill}}c@{\extracolsep{\fill}}c@{\extracolsep{\fill}}c@{\extracolsep{\fill}}c@{\extracolsep{\fill}}c@{\extracolsep{\fill}}c@{\extracolsep{\fill}}c@{\extracolsep{\fill}}c@{\extracolsep{\fill}}c@{\extracolsep{\fill}}c@{}}
  \hline \hline
& \multicolumn{8}{c}{$D_{fixed}$} & \multicolumn{11}{c}{$D_{random}$} \\ \cline{2-9} \cline{10-20}
& \multicolumn{4}{c}{$M_{fixed}$} & \multicolumn{4}{c}{$M_{random}$} & \multicolumn{7}{c}{$M_{fixed}$} & \multicolumn{4}{c}{$M_{random}$} \\ \cline{2-5} \cline{6-9} \cline{10-16} \cline{17-20}
$\bm{s}_{true} \backslash \hat{\bm{s}}$ & 1 & 2 & 3 & 4 & 1 & 2 & 3 & 4 & 1 & 2 & 3 & 4 & 5 & 6 & 7 & 1 & 2 & 3 & 4 \\ \hline
1 & 99 & 0 & 1 & 0 & 96 & 0 & 4 & 0 & 97 & 0 & 1 & 2 & 0 & 0 & 0 & 96 & 0 & 4 & 0 \\
2 & 1 & 99 & 0 & 0 & 0 & 100 & 0 & 0 & 10 & 50 & 40 & 0 & 0 & 0 & 0 & 12 & 88 & 0 & 0 \\
3 & 2 & 0 & 98 & 0 & 0 & 0 & 98 & 2 & 7 & 0 & 0 & 51 & 42 & 0 & 0 & 14 & 0 & 82 & 4 \\
4 & 0 & 0 & 0 & 100 & 0 & 0 & 0 & 100 & 0 & 0 & 0 & 0 & 0 & 56 & 44 & 0 & 0 & 2 & 98 \\
\hline
\end{tabular*}
\end{minipage}
\caption{Contingency table comparing the true group allocations $\bm{s}_{true}$ to those in the estimated optimal clusterings $\hat{\bm{s}}$. Results are shown when fitting fixed and random knot location models ($M_{fixed}$ and $M_{random}$) to fixed and random knot location datasets ($D_{fixed}$ and $D_{random}$).}
\label{tab:PEVI}
\end{table}

In contrast, when modelling the more heterogeneous (and realistic)  dataset $D_{random}$, which is more realistic in practice, the fixed knot model performs significantly worse than the random knot model. The $M_{fixed}$ model gives an ARI score which indicates poor concurrence with $\bm{s}_{true}.$ This arises as the fixed knots lead to biased estimates of each child's growth velocities $\bb_i$, which then results in a much larger estimated number of groups as the estimated growth curves are forced to be more dissimilar. True subgroup 1 has the same average velocity $(-3.0)$ across all broken stick segments. As a result, the location of the knots is not critical, and so any bias in $\bb_i$ for members in this group is relatively small. Therefore, this group can largely be identified correctly under the fixed knot model. This is not the case for the other true clusters: the individuals in these clusters tend to be split into smaller subgroups. Fitting the random knot model $M_{random}$ naturally performs well, as expected. Overall, it is clear that unless the true knot points for any growth curve are known (which will not be the case in practice) and so can be fixed in the model, the random knot location model, which allows for the heterogeneity between each individual child's growth stages, will outperform the fixed knot location model.

\section{Application: Longitudinal birth cohort in India} \label{sec:vellore}

The Healthy Birth, Growth and Development knowledge integration (HBGDki) project is an initiative supported by the Bill and Melinda Gates Foundation to combine and standardise information from various epidemiological studies into a single knowledge base \citep{jumbe2016data}. The principal objective of this project is to facilitate interdisciplinary collaboration among experts across different fields to gain insights into global child growth and development issues. The life quality of children, particularly those in low to medium income countries, can be greatly improved by the development of appropriate and timely health solutions. To date the project has amassed data sets from 192 studies, involving close to 11.5 million children and spanning 36 countries.

Our focus is on the classification of growth curves for a longitudinal study from the HBGDki project, examining the prevalence of rotavirus infections in a birth cohort in Vellore, India \citep{paul2014rotavirus}. The sample population of $N=373$ children were followed up for three years from birth and had their anthropometric measurements recorded. For the present analysis, we only analyse the HAZ scores from birth to year 1 as this is the period of fastest growth in mental development \citep{olusanya2013pattern}. We remove outliers ($\textrm{HAZ}< -6$ or $\textrm{HAZ}> 6$) based on WHO recommendations. This results in 5 to 15 observations for each child, with the first measurement taken between days 1 and 225. The time scale is represented as age in years (between 0 and 1), and the number of random growth curve knot points is specified as $K=3$. More sophisticated analyses could treat $K$ as unknown with some prior specification, which could be implemented via reversible-jump style MCMC algorithms \citep{green1995reversible,sisson05}. However, here we fix the value of $K=3$ which provides sufficient flexibility in the shape of the individual growth curves as the number of observed measurements for each child is relatively small. Further diagnostic analyses using $K>3$ did not result in noticeably different analysis outcomes (results not shown). Prior distributions on model parameters follow those in Section~\ref{sec:simulation}. The MCMC algorithm is run for 300\,000 iterations, with the first 100\,000 iterations discarded as burn in, retaining every 50th of the remaining samples. 

\begin{figure}
\centering
\includegraphics[trim=3cm 2.5cm 3cm 1cm,clip,width=\columnwidth]{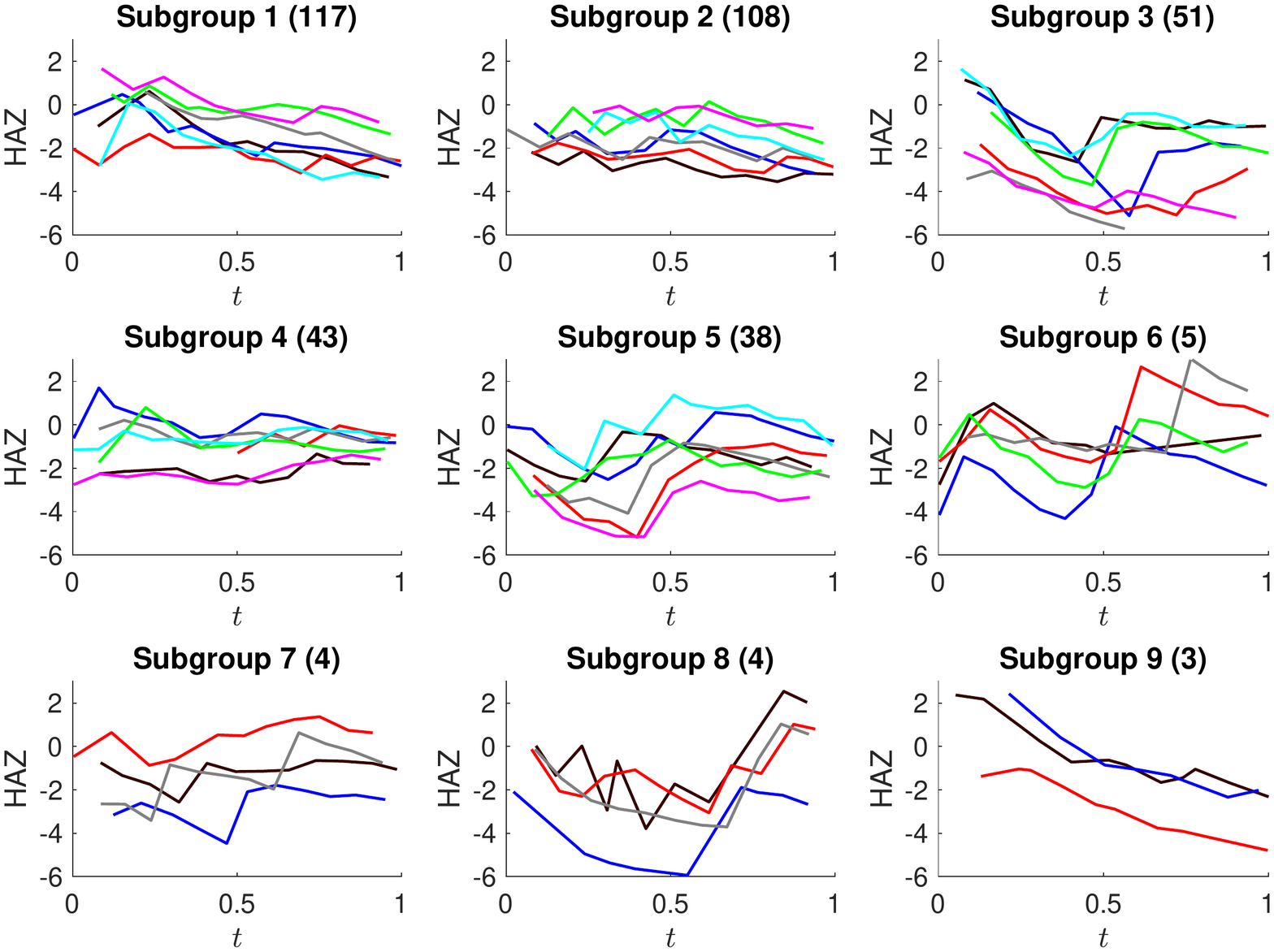}
\caption{{\small Subgroups of children from the Vellore cohort. 
Individual raw trajectories, obtained by connecting the observations with straight lines, are shown for a sample of children from each subgroup. The number of children in each subgroup is given in parentheses.}}
\label{fig:raw trajectories}
\end{figure}

\begin{figure}
\centering
\includegraphics[trim=3cm 2.5cm 3cm 1cm,clip,width=\columnwidth]{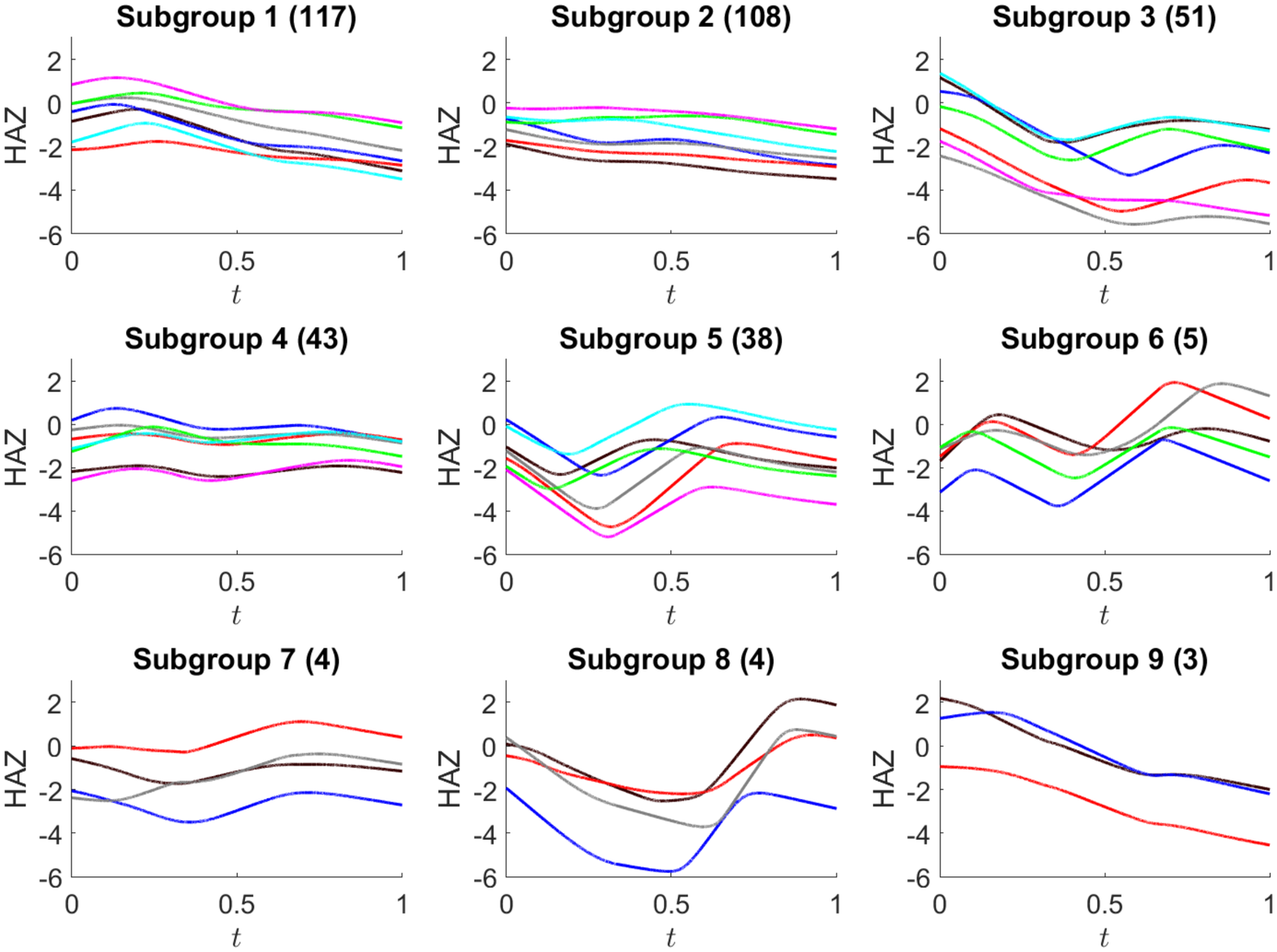}
\caption{{\small Estimated posterior mean trajectories for the same sample and groupings of children in Figure~\ref{fig:raw trajectories}.}}
\label{fig:fitted trajectories}
\end{figure}

Figure~\ref{fig:raw trajectories} shows empirical growth curves obtained from the estimated groupings of children, and Figure~\ref{fig:fitted trajectories} shows the associated fitted posterior mean growth curves. The optimal clustering $\hat{\bm{s}}$ identifies 9 unique subgroups, which coincides with the posterior mode for the number of groups in the posterior distribution for $G$. Children in the two largest subgroups show faltering growth where the distinction between them is the slight improvement in the HAZ score from birth in subgroup 1. Subgroup 3 experiences severe early stage faltering which persists for approximately 6 months, after which the HAZ scores subsequently improve. This growth pattern is also observed in subgroup 7, but the changes are milder. Subgroups 4 and 6 each alternate between growth and faltering, with the amplitude of each change differentiating between the two groups. Children in subgroups 5 and 8 exhibit a steep decline in the HAZ score before a short interval of significant recovery is observed, and which is then followed by another onset of faltered growth. The difference between these two subgroups lies in the times at which catch-up growth occurs ($t \in [0.25, 0.5]$ for subgroup 5, $t \in [0.5,0.8]$ for subgroup 8). Subgroup 9 consists of a small number of children with severe and continued faltering growth between birth and age 1.


\begin{figure}[t!]
    \centering
    \includegraphics[trim=3cm 2cm 3cm 1.5cm,clip,width=.75\columnwidth]{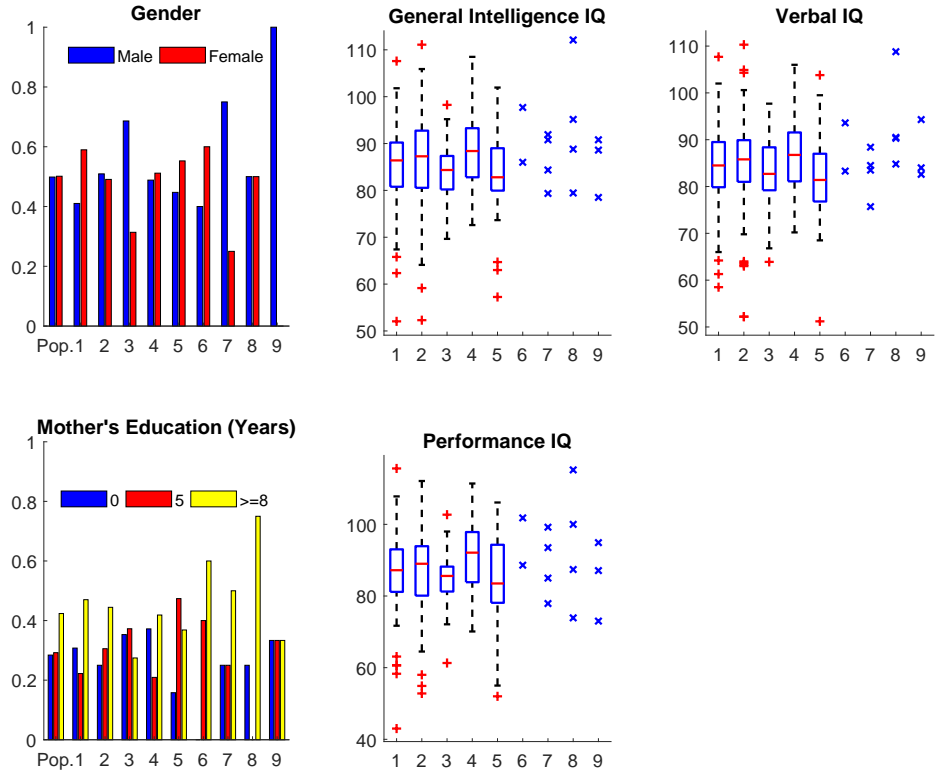}
    \caption{Bar charts illustrating the proportion of children in terms of gender and maternal education levels in different subgroups (left panels), and boxplots showing the distributions of IQ scores types (general intelligence, verbal and performance) for children in different subgroups (center and right panels). Raw data $(\times)$ for IQ scores are shown for subgroups 6--9 which have a small number of observations. Not all children are represented in each boxplot due to missing data.}
    \label{fig:analysis result}
\end{figure}

We conduct a further analysis to explore whether there is any relationship between various covariates recorded on each individual and the classification of children into the subgroups illustrated in Figure~\ref{fig:raw trajectories}. Figure~\ref{fig:analysis result} shows the results. In terms of gender composition, subgroup 1 comprises mostly females (59.0\%), whereas subgroup 3 has a disproportionately large number of males (68.6\%). These subgroups deviate significantly from the composition of the full sample which has approximately equal proportions for each gender. Mothers who received no formal education are more likely to give birth to children that exhibit the growth patterns in subgroups 3 and 4. Children are also more likely to experience severe faltering in their early childhood (subgroups 3 and 5) if they are birthed by mothers who completed 5 years (a moderate amount) of education. Moreover, these children have lower IQ scores (general intelligence, performance and verbal) compared to their peers, as indicated by the lower median scores for these tests in subgroups 3 and 5. On the other hand, children in subgroup 4, which exhibits the mildest faltering of all subgroups, have the highest median IQ scores for all tests and this is in line with the results in \cite{emond2007weight}. For children in the smaller subgroups (i.e.~subgroups 6--9), subgroups 7 and 9 are dominated by male children (75\% and 100\% respectively), while those in subgroups 6 and 8 are mostly bore by mothers who are highly educated. There are no obvious covariate patterns to account for those children who experienced severe faltering in the first year (subgroup 9). However, these conclusions are unreliable due to the small number of children in these subgroups.

\section{Conclusion} \label{sec:conclusion}

Our article proposes a new model to classify growth patterns in longitudinal child growth studies where the number of classes is not known in advance. We model the evolution of growth in terms of the HAZ scores by piecewise linear segments (i.e.~the broken-stick model) whereby an individual child's rates of growth are characterised by the slopes of these segments. Accordingly, it is plausible to use these slope parameters as a proxy for the growth pattern and so model their similarities via a mixture distribution. The classification rule is then determined by the mixture component from which the growth pattern is generated. A mixture distribution requires specifying a suitable number of components to prevent over- and under-fitting. To overcome this issue, a Bayesian non-parametric approach is adopted using the Dirichlet process prior, so that the number of mixture components is driven by the complexity of the data. 

In order to extend the flexibility of the broken-stick model, and ensure that it can be a viable model in practice given the heterogeneity inherent in observed datasets, we incorporated random knot locations into the model. The location of the knots varies between children and follows the even-numbered order statistics distribution in \cite{green1995reversible} a priori. In addition, we impose a structural restriction which ensures that there is a knot within each of several equally divided segments of the observational period. This is because we regard two growth curves, where one has the same shape as the other but lags by one period, as being different. Our simulation studies suggest that overall the random knot point model performs well: the fixed knot points model overestimates the number of components when the true data generating process has random change points due to the resulting biased estimation of the velocity vectors.

Our methodology is applied to a longitudinal study of birth cohort in Vellore, India from the Healthy Birth, Growth and Development knowledge integration (HBGDki) project funded by the Bill and Melinda Gates Foundation. Analysis of the posterior distribution indicated that there are 9 different types of growth profiles in the population. We note that the granularity of the classification can be increased if we are willing to impose stronger assumptions in the model, for example by having a shared covariance matrix across all subgroups, or by restricting the covariance matrices to be diagonal.

\section*{Acknowledgements}
The authors are all supported by the Australian Research Council through the Australian Centre of Excellence in Mathematical and Statistical Frontiers (ACEMS; CE140100049).  Lee and Ryan were previously supported by the Bill and Melinda Gates Foundation (OPP1126975).  The authors are grateful to Professor Gagandeep Kang from the Christian Medical College in Vellore India, who was the lead investigator for the study described in our Application.

\bibliographystyle{chicago}
\interlinepenalty=10000
\bibliography{biblio}
\end{document}